\documentclass[12pt,epsfig]{article}
\usepackage{graphicx}
\def\beq{\begin{equation}}
\def\eeq{\end{equation}}
\def\bea{\begin{eqnarray}}
\def\eea{\end{eqnarray}}

\begin{document}
% \begin{flushright} hep-th/9701097
% \end{flushright}
% \rightline{TIFR/TH/97-01}
% \rightline{January 1997}

\begin{center}
{\Large \bf A black potential for spin less particles
  }

\vspace{1.3cm}

{\sf  Ananya Ghatak \footnote{e-mail address: \ \ gananya04@gmail.com}, Mohammad Hasan \footnote{e-mail address: \ \ mohammadhasan786@gmail.com}
and Bhabani Prasad Mandal \footnote{e-mail address:
\ \ bhabani.mandal@gmail.com, \ \ bhabani@bhu.ac.in  }}

\bigskip

{\em $^{1,3}$Department of Physics,
Banaras Hindu University,
Varanasi-221005, INDIA. \\
$^2$ISRO Satellite Centre (ISAC),
Bangalore-560017, INDIA \\ }

\bigskip
\bigskip

\noindent {\bf Abstract}

\end{center}
We consider the most general non-Hermitian Hulthen potential to study the scattering
of spin-less relativistic particles. The conditions for CC, SS and CPA are obtained analytically
for this potential. We show that almost total absorption occurs for entire range
of incidence energy for certain parameter ranges of the potential and hence term this
as `black potential'. Time reversed of the same potential shows perfect emission for the
entire range of particle energy. We also present the classical analog of this potential in
terms of waveguide cross section.

\medskip
\vspace{1in}
\newpage

\section{Introduction}
Certain class of non-Hermitian systems with real energy eigenvalues has become the topic of 
frontier research over a decade and half because one can have fully consistent quantum theory
by restoring the Hermiticity and upholding the unitary time evolution for such system in a
modified Hilbert space \cite{ben4}-\cite{benr}. The study of non-Hermitian system has received a 
huge boost in this decade when
some of the predictions of such theories were experimentally observed in optics 
\cite{ opt1}-\cite{eqv1} and therefore 
such theories have found many applications in different branches of physics \cite{ent}-\cite{ cal}.  
Non-Hermitian theories have very rich scattering properties. Various features of scattering 
due to non-Hermitian potential like exceptional points (EPs) \cite{ep0}-\cite{ep2}, spectral singularity 
(SS) \cite{ss1}-\cite{ss3}, invisibility  \cite{aop}-\cite{inv1}, reciprocity  \cite{aop}-\cite{resc}, coherent perfect absorption (CPA) \cite{cpa00}-\cite{cpa4}
 and critical coupling (CC) \cite{cc0}-\cite{cc4} have generated huge interests during last few 
years due to their applicability and usefulness in the study of different optical systems. 
CPA or CC, the time reversal of lasing effect has become very exciting due to the discovery 
of anti-laser \cite{cpa00}-\cite{cpa011} which has number of applications in optical computer, radiology etc. 
This phenomena of perfect absorption can be observed in quantum scattering
when waves interact with the surrounding medium 
through a complex potential distribution. The event of null transmission and 
reflection in the case of unidirectional incidence is named as critical coupling. The
occurrence of CPA takes place when the waves incident from both directions on 
a potential
and then interfere with one another in such a way as to perfectly cancel each other out.
These two significant phenomena of total absorption can have more variety of consequences and 
applications in different branches of science.
However CPA and CC have so far been showed for some discrete incidence energies for certain specific complex potential \cite{dp}. Recently we have been able to find certain small ranges of incidence energies for CC and CPA in a particular system  \cite{hsn}. 

The aim of this work is to construct a 
complex potential in such a manner that it behaves almost as a perfect absorber (or a perfect emitter)
for the entire range of incidence energy.
For this purpose we consider the general one dimensional Hulthen potential written as, 
\beq
V(z)=\theta(-z)\frac{V_1}{e^{-\alpha_1 z}-Q_1}+\theta(z)\frac{V_2}{e^{\alpha_2 z}-Q_2} \ ;
\label{hul1}
\eeq
which has six independent real and/or complex parameters $V_1$, $V_2$, $Q_1$, $Q_2$, $\alpha _1$, $\alpha _2$. $\theta(z)$ the Heaviside step function is defined as,
\bea
\theta(z) &=&1 \ \ \mbox{for} \ \ z\geq 0 \ ; \nonumber \\
&=& 0 \ \ \mbox{for} \ \  z<0
\label{thet}
\eea
This potential can be parametrized in many ways for PT-symmetric (as well as non PT-symmetric) 
non-Hermitian configurations. The left and right scattering coefficients for this complex non-Hermitian potential are calculated by solving corresponding Klein-Gordon(KG) equation to capture 
 various characteristics in different parametric regime. We intend to 
focus on the non-Hermitian configurations of this potential
to study the occurrence of perfect absorption for entire range of incidence energy.

The conditions of 
null as well as super scattering are obtained analytically and are also demonstrated graphically. We find a region in the space of the parameters of this potential where it 
behaves almost\footnote
{R and T can not vanish rigorously over a range of continuous energy \cite{ref1,ref2}.} as a 
perfect absorber (or emitter) for the entire range of the incident particle energy resulting
a broadband CPA. This specific potential because of its 
absorption properties is termed as a `black 
potential'. At very low energy almost CPA is also achieved by further fine tunning the parameters within their allowed regimes.
The time 
reversed of the same potential shows the similar behavior reversely for the entire range
of the incident particle energy. The waveguide analog of this potential is also presented for 
classical realization. Physical dimension of the waveguide and operating frequency are calculated.

Different aspects of scattering and absorption in the Non-Hermitian potential are discussed 
in Sec. 2. Sec. 2.1 is devoted for low energy scattering and critical coupling.
Bidirectional absorption is discussed in Sec. 2.2. Waveguide 
analog for this potential is demonstrated in Sec. 3. Sec. 4 is kept for conclusions and 
discussions. Necessary mathematical details are provided in Appendix.

\section{Scattering and absorption in the Non-Hermitian Hulthen potential}

The most general non-Hermitian structure of the Hulthen potential in Eq. \ref{hul1} is 
obtained by complexifying all the  parameters as,
\beq
V_{1,2}\rightarrow v_{1,2} e^{i \beta_{1,2}} \ ; \ \ \alpha_{1,2}\rightarrow  a_{1,2} e^{i \gamma_{1,2}} \ ;
\ \ Q_{1,2}\rightarrow q_{1,2} e^{i \phi_{1,2}} \ .
\label{para}
\eeq
where $v_{1}$, $v_{2}$, $a_{1}$, $a_{2}$, $q_{1}$, $q_{2}$ are all real numbers and assumed to be positive 
definite for convenience and the arguments $ \beta_{1}$, $ \beta_{2}$,  $\gamma_{1}$, $ 
\gamma_{2}$, $ \phi_{1}$, $ \phi_{2}$ are all taken between $0$ to $ 2\pi$. $v_{1,2}$, 
$a_{1,2}$ have dimension of energy and length inverse respectively where as $q_{1,2}$,
$ \beta_{1,2}$,  $\gamma_{1,2}$ and $ \phi_{1,2}$ are dimensionless.
We solve the 1-D KG equation,
\beq
\frac{d^2\phi(z)}{dz^2}+\Big\{\frac{[E-V(z)]^2}{c^2\hbar^2}-\frac{m^2c^2}{\hbar^2}\Big\}
\phi(z)=0
\label{kg}
\eeq
for non-Hermitian potential in Eq. \ref{hul1} with complex parameters defined
in Eq. \ref{para} for $z<0$ and $z>0$ by following Ref. \cite{hul1}. The wave functions and 
their asymptotic behaviors are explicitly given in Appendix. For convenience we adopt
natural unit ($\hbar=c=1$ ) and chose the mass ($m$) of the incident particle as $ 1 MeV$,
$v_{1,2}$ and $a _{1,2}$ in $MeV$ and $z$ in $MeV^{-1}$ for physical realization. 
$MeV^{-1}$ can be converted in  nanometer (nm) using the conversion relation
$1 MeV^{-1}=1.97\times 10^{-4} nm$ for practical purposes.

The scattering amplitudes for left and right incidence are calculated by
demanding the continuity of the wave functions and their derivatives at $z=0$. 
The scattering amplitudes for left incident case are written in the compact form as,
\begin{equation}
\label{rl}
r_{l}=Q_{1}^{2\mu_{1}} \frac{\Big[-F_1F_2(N_1+N_2)+c_1F_1^{'}F_2+c_3F_2^{'}F_1\Big]}
{2MF_4F_2+F_4F_2(N_1+N_2)-c_2F_4^{'}F_2-c_3F_2^{'}F_4}\equiv Q_1^{2\mu_1} 
\tilde{F_{rl}} \ ;
\label{rl}
\end{equation}

\begin{equation}
\label{tl}
t_{l}=Q_{1}^{2\mu_{1}} \frac{\Big[2MF_1F_4+c_1F_1^{'}F_4-c_2F_4^{'}F_1\Big]}
{2MF_4F_2+F_4F_2(N_1+N_2)-c_2F_4^{'}F_2-c_3F_2^{'}F_4} \left(\frac{1}{u}\right )\equiv \frac{Q_1^{2\mu_1}}{u}\tilde{F_{tl}} \ ;
\label{tl}
\end{equation}
where,
\beq
\mu_1 =ik/\alpha_1 \ \ ; \ \ \mu_2 =ik/\alpha_2 \ \ ; \ \ k=\sqrt{E^2-1} \ ;
\label{mu}
\eeq
\beq
N_1=\frac{Q_1\alpha_1 \lambda_1 }{1-Q_1} \ ; \ 
N_2=\frac{Q_2\alpha_2 \lambda_2 }{1-Q_2} \ ; 
\label{nn}
\eeq
\beq
 u=\frac{Q_2^{-\mu_2}(1-Q_2)^{\lambda_2}}{Q_1^{-\mu_1}(1-Q_1)^{\lambda_1}} \ ;
 \eeq
 \bea
c_1&=&\frac{(\mu_1 +\lambda_1 )^2-\nu_1 ^2}{1+2\mu_1 }(Q_1\alpha_1 ) \ \ ;  \ \ c_2=\frac{(-\mu_1 +\lambda_1 )^2-\nu_1 ^2}{1-2\mu_1 
}(Q_1\alpha_1 ) \ ; \nonumber \\ 
c_3&=&\frac{(-\mu_2 +\lambda_2 )^2-{\nu_2} ^2}{1-2\mu_2 }(Q_2\alpha_2 ) \ \ ; \ \ c_4=\frac{(\mu_2 
+\lambda_2 )^2-{\nu_2} ^2}{1+2\mu_2 }(Q_2\alpha_2 ) \ ;
\label{css}
\eea
with
\beq
\lambda_1 =1/2+1/2\sqrt{1-\Big(\frac{2V_1}{\alpha_1 
Q_1}\Big)^2} \ \ ; \ \lambda_2 =1/2+1/2\sqrt{1-\Big(\frac{2V_2}{\alpha_2
Q_2}\Big)^2} \ ;
\label{l1l2}
\eeq
\beq
\nu_1 =\sqrt{\mu_1 ^2+\lambda_1^2-\lambda_1 -\frac{2EV_1}{\alpha_1 ^2 Q_1}} \ \ ; \ \ \nu_2 
=\sqrt{\mu_2^2+\lambda_2^2-\lambda_2 -\frac{2EV_2}{\alpha_2^2 Q_2}} \ .
\label{nu1nu2}
\eeq

The functions $\tilde{F_{rl}}$ and $ \tilde{F_{tl}}$ in reflection
and transmission amplitudes in Eqs. \ref{rl} and \ref{tl} contain the contributions from
hyper-geometric functions (given in Eqs. \ref{hyp1}-\ref{hyp4} in Appendix). 
At the very outset we concentrate on the CC for left incidence (i.e.  $T_l\equiv \ \mid t_l\mid ^2 \rightarrow 0$ and $R_l\equiv \ \mid r_l\mid ^2\rightarrow 0$) by looking
at the behavior of $r_l$ and $t_l$ for different incidence energies. We first look at the 
behavior of the terms  $Q_{1}^{2\mu_{1}}$ in $r_l$  and $Q_{1}^{2\mu_{1}}\frac{1}
{u}$ in $t_l$ which involve no hyper-geometric functions. 
$Q_{1}^{2\mu_{1}}$ is written in terms of the parameters in Eq. \ref{para} as,
\begin{equation}
Q_{1}^{2\mu_{1}}=(q_{1}e^{i\phi_{1}})^{2\frac{k}{a_{1}} e^{i( \frac{\pi}{2} -\gamma_{1})} } \ .
\label{qmu}
\end{equation}
Using an identity 
\beq
|(A e^{ia})^{B e^{ib}}|= e^{B (\cos{b} \ln{A}- a\sin{b})}, \ \ \mbox{where $A, B, a, b$ are all real,}
\label{ind}
\eeq 
we further re-express Eq. \ref{qmu} as,
\begin{equation}
\label{condition1_q1}
\vert Q_{1}^{2\mu_{1}} \vert= e^{\frac{2k}{a_{1}} (\sin{\gamma_{1} \ln{q_{1}}}-\phi_{1}\cos{\gamma_{1}}) }\equiv  e^{\frac{2k }{a_{1}}\xi_{1}}
\label{q1m}
\end{equation}
with \beq
\xi_1=\sin{\gamma_{1} \ln{q_{1}}}-\phi_{1}\cos{\gamma_{1}}
\eeq

It is clear from Eq. \ref{q1m} that $\vert Q_{1}^{2\mu_{1}} \vert$ can play an important 
role as it contributes exponentially with increasing energy. The contribution of this term be
enhanced further for small $a_{1}$. As shown in Fig. 1(a) this term has huge contribution to $r_l$ 
with increasing energy for smaller values of $a_1$. In 
comparison to this the behavior of $\tilde{F_{rl}}$ (for $\vert Q_{1}\vert<1$) shows saturating nature
with increasing energy (Fig. 1(b)). Therefore that $\vert Q_{1}^{2\mu_{1}} \vert$ 
is the most dominating term in $r_l$ with increasing energy and is responsible for super or null reflectivity. The 
occurrence of null reflectivity depends upon the sign 
of the term $\xi_{1}$ in the exponential of Eq. \ref{q1m}. We define the critical value of the parameter $\gamma_{1}$ as,
\beq
\gamma_{1}^c=\tan^{-1}\frac{\phi_{1}}{\ln{q_{1}}}
\eeq
For $\gamma_{1}>\gamma_{1}^c$, $\xi_1$ is positive and $R_l=\vert r_l\vert ^2$ is 
going to very high values with increasing energy (Fig. 1(c)).  \\

\includegraphics[scale=1.6]{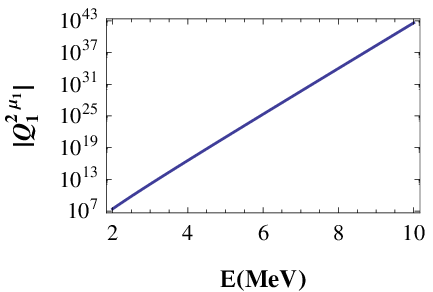} \ a \ \includegraphics[scale=1.6]{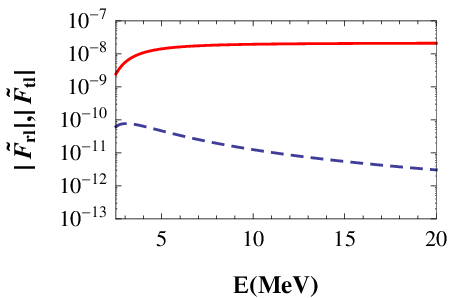} \ b \ \\

\includegraphics[scale=1.6]{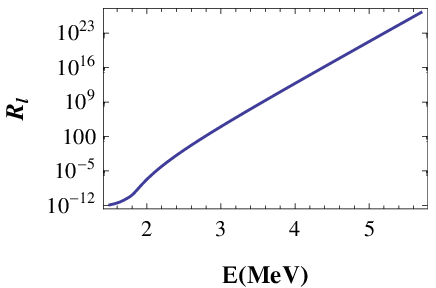} \ c \ \includegraphics[scale=1.6]{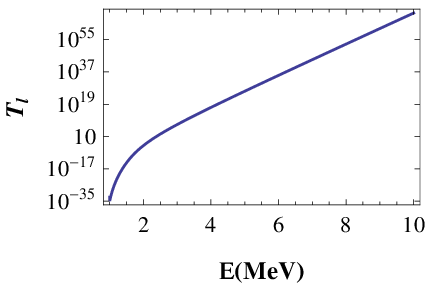} \ d \ \\

{\it Fig. 1: The behavior of $\vert Q_{1}^{2\mu_{1}} \vert$ in 1(a) and $\left|\tilde{F_{rl}}
\right|$ (dotted line),  
$\left|\tilde{F_{tl}}\right|$ (solid line) in 1(b) are shown for 
the choice of the parameters as $q_1 = q_2 = 0.4, v_1= v_2 = 1 MeV, a_1 = a_2 = 0.1 MeV, 
\phi_{1} = 
\phi_{2} = 1.3, \gamma_{1} = \gamma_{2} = 2.5, \beta_{1} =\beta_{2} = 0.2,$ leading 
$\gamma_{1}>\gamma_{1}^c$. $R_l$ and $T_l$ with the same set of parameters are plotted in 1 (c) and 1 (d) respectively.} \\

On the other hand for 
$\gamma_{1}<\gamma_{1}^c$ we have almost null reflectivity ($R_l\sim 10^{-24}$ and less)
for the entire range of energy (Fig. 2(a))
except at very low energy. The scattering at very low energy is different in 
nature and has been
dealt with separately in the next section. \\

Similarly for transmission amplitude $t_l$ the dominant factor (hyper-geometric function independent part) is written using the expression of $u$ in Eq. \ref{nn} as,
\begin{equation}
Q_{1}^{2\mu_{1}} \frac{1}{u} = Q_{1}^{\mu_{1}} Q_{2}^{\mu_{2}}  \frac{ (1-Q_{1})^{\lambda_
{1}}} {(1-Q_{2})^{\lambda_{2}}}\equiv f_1f_2
\end{equation}
(where $\mu_{1,2}$ and $\lambda_{1,2}$ are given in Eqs. \ref{mu}, \ref{l1l2}) with
\beq
f_{1}=Q_{1}^{\mu_{1}} Q_{2}^{\mu_{2}} \ ; \ \ f_{2}= \frac{ (1-Q_{1} )^{\lambda_{1}}} {(1-Q_{2} )^{\lambda_{2}}}
\label{f1f2}
\eeq
so that
\begin{equation}
\label{modq1u}
\vert Q_{1}^{2\mu_{1}} \frac{1}{u} \vert = \vert f_{1}\vert \vert f_{2}\vert
\end{equation}
$\vert f_{2}\vert$ depends only on the potential parameters and independent of $k$.
The energy dependence of $\vert Q_{1}^{2\mu_{1}} \frac{1}{u} \vert$ is only through $\vert f_1
\vert$, which is re-expressed (again by using the identity in Eq. \ref{ind}) as, 
\begin{equation}
\vert f_{1} \vert= e^{ k[\frac{1}{a_{1}} (\sin{\gamma_{1} \ln{q_{1}}}-\phi_{1}\cos{\gamma_{1}})  +\frac{1}{a_{2}} (\sin{\gamma_{2} \ln{q_{2}}}-\phi_{2}\cos{\gamma_{2}}) ]}
\equiv e^{k[\frac{\xi_1}{a _1}+\frac{\xi_2}{a _2}]}
\label{exp_f1}
\end{equation}
The variation of $\tilde{F_{tl}}$ in Eq. \ref{tl} is not dominating 
as shown in the Fig. 1(b). The behavior of $T_l=\vert t_l\vert ^2$ depends on the 
overall sign of the exponential in $\vert f_1\vert$. We already have the reflection-less condition 
for $\xi_1<0$, so we discuss the possible conditions of CC in high energy ($a_{1,2}<k$) range as follows. \\

{\bf Case 1:}  ${\bf \xi_2<0 \ ; \ \xi_1<0 \ ; \ a_{1,2}<k ;}$

Since $\xi_1, \xi_2$ both are negative we have total absorption with increasing energy of 
left incident particle. The critical values of $\gamma_{1,2}$
are written in terms of the other parameters as,
\beq
\gamma_{1,2}^c=\tan^{-1}\frac{\phi_{1,2}}{\ln{q_{1,2}}};
\label{gam}
\eeq
Thus we have total absorption for
entire range of both left and right incident waves for $\gamma_{1,2}<\gamma_{1,2}^c$. 
For symmetric Hulthen potential (i.e. when 
$\alpha _{1}=\alpha_{2}, Q_1=Q_2$ and
$V_1=V_2$) if we consider a special case of $q_{1, 2}=1$, then CC occurs when $\gamma_{1}$ lies between $0 
\rightarrow  \pi/2 \ \mbox{or} \ \frac{3\pi}{2} \rightarrow  2\pi$ with any arbitrary value of
 $\phi_1$ between $0$ to $2 \pi$. \\
 
\vspace{1.4in}

{\bf Case 2:} $\bf{ \xi_2>0\ ; \ \xi_1<0\ ; \ a_{1,2}<k ;}$

The condition of 
critical coupling is achieved if and only if $\left(-\frac{\mid\xi_1\mid}{a _1}+\frac{\mid\xi_2\mid}{a _2}\right)<0$ which suggests,
\bea
\mid\xi_1\mid &>& \frac{a _1}{a _2}\mid\xi_2\mid \ \ \ \ \mbox{i.e.} \ , \nonumber \\
\phi_{1}\cos{\gamma_{1}} +\frac{a _1}{a _2}\phi_{2}\cos{\gamma_{2}}&>& \frac{a _1}{a _2} \sin{\gamma_{2} \ln{q_{2}}}+\sin{\gamma_{1} \ln{q_{1}}}
\label{gam2}
\eea
Eq. \ref{gam2} is realized in a simpler way if we take a 
specific choice of some of the parameters as $q_1=q_2, \phi_1=\phi_2$ and $a _1=a 
_2$. For this choice we get a upper bound for $(\gamma_1+\gamma_2)$ from Eq. \ref
{gam2} to have CC as,
\beq
(\gamma_1+\gamma_2)^c=2 \tan^{-1}\left(\phi_1/ \ln q_1\right).
\eeq \\
From the above two cases we note that when $\gamma_{1,2}>\gamma_{1,2}^c$ [or 
$(\gamma_1+\gamma_2)>(\gamma_1+\gamma_2)^c$] we get resonances in transmissivity 
for all ranges of incidence energy. Fig. 1(d) shows the very high values of $T_l=\vert 
t_l\vert^2$ as incident energy increases.
On the other hand almost null transmissivity ($T_l\sim  10^{-36}$)
for left incidence (Fig. 2(b)) occurs for almost all the values of incidence energy (except 
low energy) when $\gamma_{1,2}<\gamma_{1,2}^c$ or $(\gamma_1+\gamma_2)<(\gamma_1+\gamma_2)^c$. \\

\includegraphics[scale=1.3]{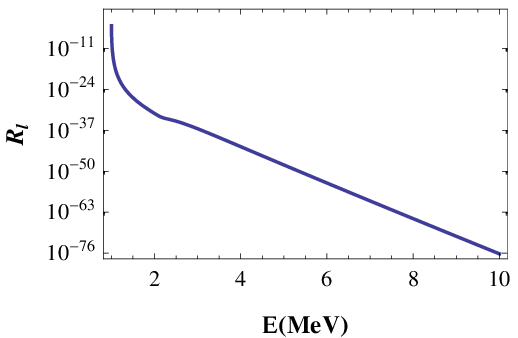} \ a \ \ \includegraphics[scale=1.3]{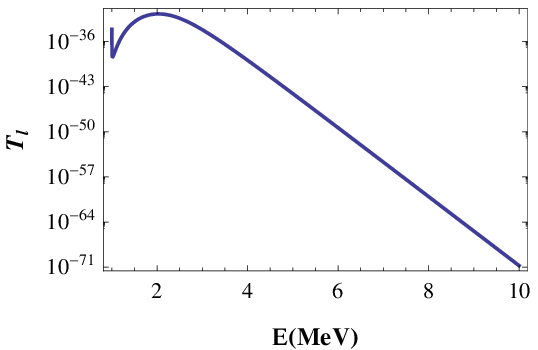} \ b \ \\

{\it Fig. 2: $R_l$ (in 2(a)) and $T_l$ (in 2(b)) are plotted with incidence energy 
by changing $\phi_{1,2}( =0.3) $ and keeping other parameters same as Fig.1 for $\gamma_{1}<\gamma_{1}^c$ . Almost null transmissivity and reflectivity for left
incidence are seen
($T_l\sim  10^{-36}$ or less and $R_l\sim 10^{-24}$ or less) at all energies except at low incidence energy.} \\

Thus we have almost null scattering for unidirectional incidence 
occurs for almost all the values of incidence energy (except at low energy) subjected
to the conditions $a_{1,2}<k$, $\gamma_{1,2}<\gamma_{1,2}^c$ or 
$(\gamma_1+\gamma_2)<(\gamma_1+\gamma_2)^c$ when both the scattering amplitudes $R_l$ and $T_l$ are almost 
vanishing (as shown in Fig. 2). However when $\gamma_{1,2}>\gamma_{1,2}^c$ [or 
$(\gamma_1+\gamma_2)>(\gamma_1+\gamma_2)^c$] we get resonances in scattering amplitudes for  
all ranges of incidence energy (Fig. 1(c,d)). Surprisingly this situation is 
different from the time reversed case of CC. The same characteristics are also seen for the 
right incidence case. Real and imaginary parts of the symmetric Hulthen potential 
 is plotted in Fig. 3(a) keeping the parameters same as in Fig. 2. The imaginary part of the 
potential (dashed curve in Fig. 3(a)) is 
responsible for the unidirectional total absorption of almost all the incident energies 
except the low energy. Fig. 3(b,c,d) are relevant to the discussion of low energy scattering in the next section.

\includegraphics[scale=1.4]{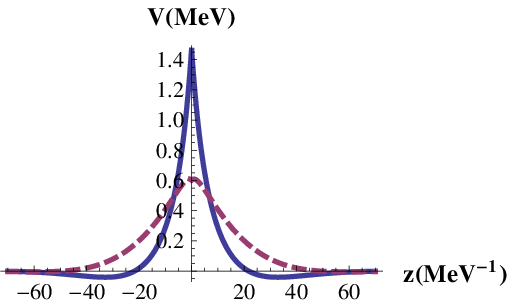} a \ \includegraphics[scale=1.2]{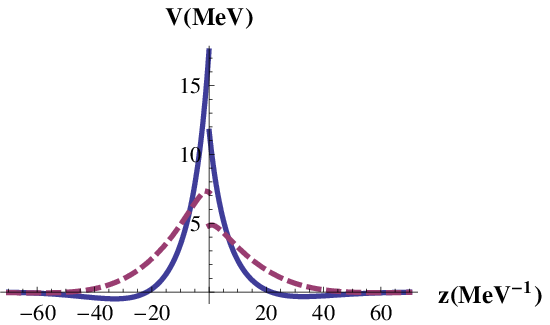} b \\

\includegraphics[scale=1.3]{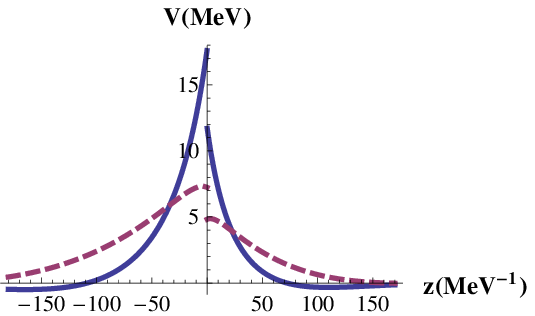} c \ \includegraphics[scale=1.2]{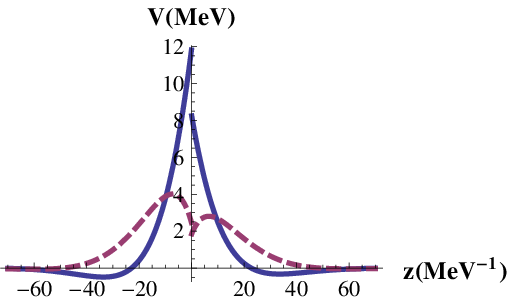} d \\

{\it Fig. 3: The real part (solid lines) and imaginary part (dashed lines) of symmetric (3(a))
as well as asymmetric (3(b,c,d)) non-Hermitian Hulthen 
potential are plotted in the parametric regime of CPA. Fig 3(a) shows the symmetric potential in the same parametric regime of Fig. 2. Fig 3(b) is plotted by changing the values of  
$v_{1} (= 12 MeV)$ and $v_{2} (= 8 MeV)$. In Fig 3(c) $q_{1,2}$ are changed to $ q_{1} = 
0.01$ and $ q_{2} = 0.06$. Fig 3(d) is for different $a_{1} (= 0.02 MeV^{-1})$ and $a_{2} (= 0.03 MeV^{-1})$.}

\subsection{Low energy scattering} 
In the low energy region the phenomena of total absorption may not occur even though 
the condition of $\gamma_{1,2}<\gamma_{1,2}^c$ [or 
$(\gamma_1+\gamma_2)<(\gamma_1+\gamma_2)^c$] is satisfied. This is also clear from the
Figs. 1 and 2. At low energy region,
$k\simeq a_{1,2}$ and therefore the
contributions of the exponential terms in Eqs. \ref{q1m} and \ref{exp_f1} are not 
dominating. However we obtain perfect absorption even at low energy region by tunning the
parameters further within their ranges and by studying the behavior of the other terms
 in $r_l$ and $t_l$. 
In the following subsections we first study the terms with hyper-geometric functions 
to see if they have important impact 
on deciding the extreme behavior of $r_l$ and $t_l$ in this region.
We further study the impact of the energy independent term $f_2$ present in $t_l$
in this low energy region.

\subsubsection{The terms $\tilde{F_{tl}}$ and $\tilde{F_{rl}}$}

Each hyper-geometric functions (see Eq. \ref{hyp1} -\ref{hyp4} for the explicit expressions) in $\tilde{F_{tl}}$ and $\tilde{F_{rl}}$ converge with 
respect to energy to an asymptotic value for $\vert Q_{1,2}\vert=q_{1,2}<1$. But in the low energy region
they have variety of behaviors depending on the different values of the parameters. The 
terms $\lambda_{1,2}$ and $\nu_{1,2}$ (given in Eqs. \ref{l1l2} and \ref{nu1nu2}) in all the hyper-geometric functions
have the following limiting cases in the
low energy region when $k$ is comparable to $a_{1,2}$,
\beq
\hspace{-3.5in}\mbox{(I) \ \ for}  \ \left | \frac{V_{1,2}}{\alpha_{1,2}Q_{1,2}}\right |>>1 \ :
\label{e2}
\eeq
we can write from Eqs. \ref{l1l2} and \ref{nu1nu2},
\beq
 \ \ \vert\lambda_{1,2}\vert\approx \left |\frac{V_{1,2}}{\alpha_{1,2}Q_{1,2}}\right | \ ; \
 \vert\nu_{1,2}\vert\approx \left |\frac{V_{1,2}}{\alpha_{1,2}Q_{1,2}}\right | \ ,
\label{ee2}
\eeq
as in this low energy region $\vert \mu _{1,2}\vert =\vert \frac{i k}{\alpha  _{1,2}}\vert\sim 1$ and we see from Eq. \ref{nu1nu2} that
$\Big |\mu_{1,2} ^2-\lambda_{1,2} -\frac{2EV_{1,2}}{\alpha_{1,2} ^2 q_{1,2}}\Big |\sim \vert\lambda_{1,2}\vert $ and
$\vert\lambda_{1,2}^2\vert >> \vert\lambda_{1,2}\vert $.
\beq
\hspace{-3.5in} \mbox{(II) \ \ for}  \ \left | \frac{V_{1,2}}{\alpha_{1,2}Q_{1,2}}\right |<<1 \ : 
\eeq
from Eqs. \ref{l1l2} and \ref{nu1nu2} we see that
\beq
 \lambda_{1,2}\approx 1 \ \ ; \ \ \nu_{1,2}\approx\sqrt{\mu_{1,2}^2- \frac{2 V_{1,2}E}{\alpha_{1,2}^2Q_{1,2}}}\approx \mu_{1,2} \ .
\label{e1}
\eeq
Now for
simplicity we consider a symmetric complex Hulthen potential ($\alpha_{1}=\alpha_{2}, Q_1=Q_2,
V_1=V_2$) for case-I. Putting the values of $\lambda_{1}, \nu_{1}$ from  
Eq. (\ref{ee2}) in the Eqs. (\ref{nn}) and (\ref{css}) we obtain,
\beq
N_1=\frac{iV_1}{1-Q_1}=N_2 \ ; \ c_1=c_4=\frac{2i \mu_1V_1}{1+2\mu_1}= -c_2=-c_3 \ .
\label{nc}
\eeq
For this symmetric case we see from Eqs. (\ref{hyp2}) and (\ref{hyp4}) that $F_2=F_4$ and
$F_2^{'}=F_4^{'}$, therefore in this limit $\tilde{F_{rl}}$ is written by using Eq. (\ref{nc}) as,
\beq
\left|\tilde{F_{rl}}\right |=\frac{\left|-F_1F_4(\frac{2V_1}{1-Q_1})+(\frac{2 \mu_1V_1}{1+2\mu_1})F_1^{'}F_4-(\frac{2\mu_1 V_1}{1+2\mu_1})F_4^{'}F_1\right |}
{\left| 2Me^{\frac{-i\pi}{2}}F_4^2+F_4^2(\frac{2V_1}{1-Q_1})+(\frac{2 \mu_1 V_1}{1+2\mu_1})F_4^{'}F_4+(\frac{2 \mu_1 V_1}{1+2\mu_1})F_4^{'}F_4 \right |}
\label{frllm}
\eeq
where $F_{i=1,2,3,4}$ (and ${F^{'}}_{i=1,2,3,4}$) are given in Appendix.
Further in this limiting case, it can be shown by using the properties of hyper-geometric 
function \cite{hbook} that $|F_1|\approx |F_4|$ and $|F_1^{'}|\approx |F_4^{'}|$ as 
$|\nu_1|$ and $|\lambda_1|$ are being much larger comparable to $|\mu_1|$ in low energy region. Hence from
 Eq. \ref{frllm} it is easily seen that the denumerator is 
 always larger than the numerator. Therefore for convenient values of $v_1$ 
and/or $a_1$, the reflection coefficient vanishes as $\left|\tilde{F_{rl}}\right |\rightarrow 0$ 
in the low energy region. These have been demonstrated in Figs. 4(a, b). 

Similarly for transmission coefficient
in this energy region and for a symmetric potential case the term $\tilde{F_{tl}}$ is written as,
\beq
\left|\tilde{F_{tl}}\right |=\frac{\left|2 M e^{\frac{-i\pi}{2}}F_1F_4+(\frac{2 \mu_1V_1}{1+2\mu_1})F_1^{'}F_4+(\frac{2\mu_1 V_1}{1+2\mu_1})F_4^{'}F_1\right |}
{\left| 2Me^{\frac{-i\pi}{2}}F_4^2+F_4^2(\frac{2V_1}{1-Q_1})+(\frac{2 \mu_1V_1}{1+2\mu_1})F_4^{'}F_4+(\frac{2\mu_1 V_1}{1+2\mu_1})F_4^{'}F_4 \right |}
\label{ftllm}
\eeq
From this equation we see that $\left|\tilde{F_{tl}}\right |$ goes to zero even faster than 
$\left|\tilde{F_{rl}}\right |$ resulting $T_l\rightarrow 0$ in the low energy region (Figs. 
4(a, b)). On the other hand from case II, we do not get such parametric regime for
which perfect absorption is achieved. 
Thus in this low energy region these hyper-geometric functions play important roles
over the previously discussed exponential terms and lead to total absorption. 
We further study the total absorptivity ($A_l=1-R_l-T_l$) for left incidence with
different values of $v_1$ following the above discussion. $A_l$ has been plotted in Figs. 4(c) and 4(d)
for two different values of $v_1$ to show the total absorption at low energy.
 \\

\includegraphics[scale=1.3]{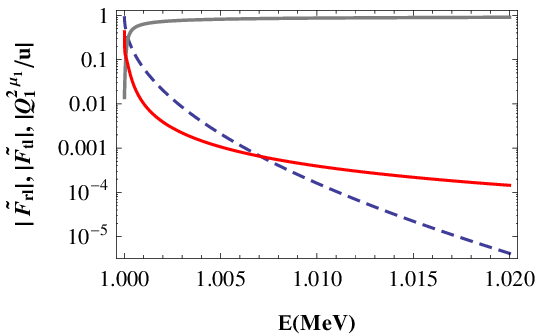} \ a \ \includegraphics[scale=1.3]{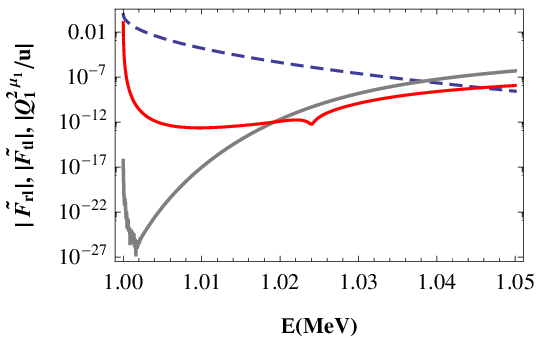} \ b \\

\includegraphics[scale=0.96]{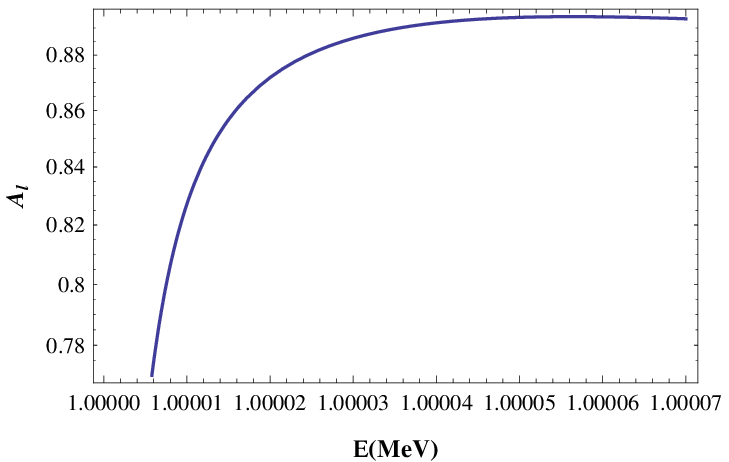} \ c  \ \includegraphics[scale=0.96]{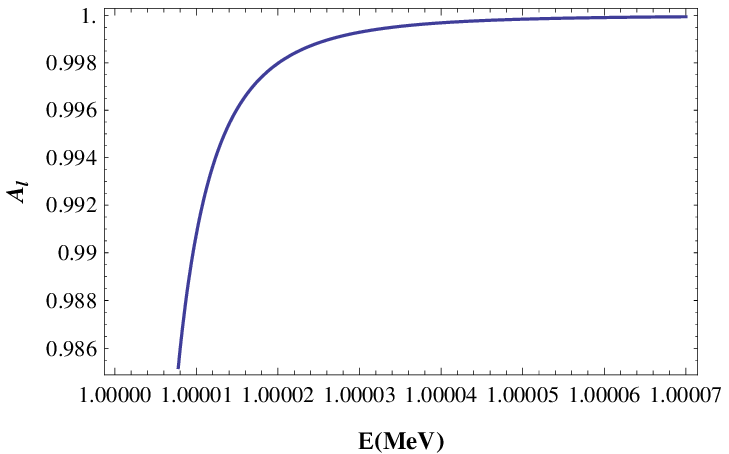} \ d \\

{\it Fig. 4:  4 (a) and 4 (b) are showing the natures $\left|\tilde{F_{rl}}\right |$ (solid 
red line) , 
$\left|\tilde{F_{tl}}\right |$ (solid gray line) and $\left |Q_{1}^{2\mu_{1}} \frac{1}{u} 
\right |$ (dashed line) in low energy
region for two different values of $v_{1,2}$,  $v_{1,2}=.0001 MeV$ and $v_{1,2}=.02 MeV$ respectively and $a_{1,2}=.01 MeV^{-1}$. 4 (c) (with $v_{1,2}=.0001 MeV$) and 4 (d) (with $v_{1,2}=.02 MeV$) show the total absorption in low energy ranges. }
 \\
 
The real and imaginary parts of the non-Hermitian asymmetric Hulthen potentials are
plotted in Fig. 3(b,c,d) with different parameters for which almost perfect absorption occurs at high
energy as well as in the low energy limit (subjected to the condition in Eq. \ref{e2}). Fig. 
3 also clarifies the contributions of the parameters ($v_{1,2}, a_{1,2}, q_{1,2} $) to the real and imaginary parts of the potential.

\subsubsection{Effect of $f_2$ term}
Now we show that even the energy independent term $f_2$ plays certain role in deciding
total absorption at low energy.
The expression of $f_2$ in Eq. \ref{f1f2} is rewritten as, 
\begin{equation}
f_{2}= \frac{\left( \sqrt{1-Q_{1} }\right)^{2\lambda_{1}}} {\left(\sqrt{1-Q_{2}}\right)^{2\lambda_{2}}}= \frac{ (\rho_{1}e^{i\tau_{1}})^{2\lambda_{1}}  } {(\rho_{2}e^{i\tau_{2}})^{2\lambda_
{2}}}
\label{f2}
\end{equation}
where 
\beq
\label{rho1}
\rho_{1,2}=(1+q_{1,2}^{2}-2q_{1,2}\cos{\phi_{1,2}})^{\frac{1}{4}} \ ; \ \  \tau_
{1,2}=\frac{1}{2} \tan^{-1} \Big[ \frac{q_{1,2}\sin{\phi_{1,2}} }{ q_{1,2}\cos{\phi_{1,2}} -1 
} \Big]
\eeq
We have already seen the criterion for total absorption for low energy region in the previous section as
$\Big |\frac{V_{1,2}}{\alpha _{1,2}Q_{1,2}}\Big |>>1$.
From Eq. \ref{ee2} the terms $\lambda_{1}$ and $\lambda_{2}$ are rewritten for the parameters given in Eq. \ref{para} as, 
\begin{equation}
\lambda_{1,2}=\left(\frac{v_{1,2}}{a_{1,2}q_{1,2}}\right) e^{i(\frac{\pi}{2}+\delta_
{1,2})}\equiv l_{1,2}e^{i(\frac{\pi}{2}+\delta_
{1,2})} \ \ ; \ \mbox{with} \ \delta_{1,2}=\beta_{1,2}-\gamma _{1,2}-\phi_{1,2}
\end{equation} 
where $l_{1,2}=\frac{v_{1,2}}{a_{1,2}q_{1,2}} $.
Using the identity in Eq. \ref{ind} we write Eq. \ref{f2} after a small trigonometric simplification as, 
\begin{equation}
\vert f_{2}\vert= \frac{e^{- 2\lambda_{a} [\sin{\delta_{1}} \ln{\rho_{1}} +\tau_{1} \cos{
\delta_{1}}]}}{ e^{- 2\lambda_{b} [\sin{\delta_{2}} \ln{\rho_{2}} +\tau_{2} \cos{\delta_{2}}]}}
\equiv e^\Gamma
\end{equation}
where 
\begin{equation}
\Gamma= - 2\lambda_{a} [\sin{\delta_{1}} \ln{\rho_{1}} +\tau_{1} \cos{\delta_{1}}] + 2\lambda_
{b} [\sin{\delta_{2}} \ln{\rho_{2}} +\tau_{2} \cos{\delta_{2}}] 
\end{equation}
$l_{1,2}$ are large for the conditions of CC at low incidence 
energy. Thus $f_2$ contributes substantially depending on the value of $\Gamma$. The sign of $\Gamma$ will
decide whether the transmissivity will be zero or diverging in these low energy regions.
However $f_2$ has no role in case of symmetric Hulthen potential for which $\Gamma =0$.

\subsection{Bidirectional perfect absorption}
We have discussed so far about the situation of total absorption when particles are coming 
only from one direction. Now we wish to explore the possibility of total absorption for all 
energies when particles are coming from both the directions. We first re-express the 
scattering amplitudes in Eqs. \ref{rla}-\ref{tta} in the Appendix by introducing $G_i$ 
($i=1,2,3,4$) as,
\begin{equation}
r_{l}= Q_{1}^{2\mu_{1}} \frac{G_1}{D} \ ; \
t_{l}= \frac{Q_{1}^{2\mu_{1}}}{u} \frac{G_2}{D} 
\label{ll}
\end{equation} 
\begin{equation}
r_{r}= Q_{2}^{2\mu_{2}} \frac{G_3}{D} \ ; \
t_{r}= Q_{2}^{2\mu_{2}} u \frac{G_4}{D} 
\label{rr}
\end{equation} 
$D$ is defined in the Eq. \ref{d} in appendix.
For CPA the determinant of scattering matrix should vanish, i.e. $\mid t_{l}t_{r}-r_{l}r_{r}\mid=0$. Using Eqs. \ref{ll} and \ref{rr} the condition
for CPA is expressed as,
\begin{equation}
Q_{1}^{2\mu_{1}}Q_{2}^{2\mu_{2}}(G_1G_2-G_3G_4)=0 
\label{cpa}
\end{equation}
The above condition is satisfied if $G_1G_2=G_3G_4$, which describes the occurrence of coherent 
perfect absorption at discrete incidence energies of the waves coming from both the directions. The other conditions
of total absorption in Eq. {\ref{cpa}} show critical coupling for which $Q_{1}^{2\mu_{1}}=0$ 
(CC for left incidence) or $Q_{2}^{2\mu_{2}}=0$ (CC for right incidence). However it is 
possible to find a parametric regime of the potential for which 
$Q_{1}^{2\mu_{1}}, Q_{2}^{2\mu_{2}}$ are extremely small  for almost all incident energy
leading to broadband CPA for bidirectional incidence. 
The values of the parameters of the potential to satisfy the criteria $Q_{1}^{2\mu_{1}}\simeq 0 \ \mbox{and} \ Q_{2}^{2\mu_{2}}\simeq 0$
are already elaborated in the previous sections. Thus 
we have possible parametric regions for which the non-Hermitian Hulthen potential (in Eq. \ref
{hul1}) with its real and imaginary parts behaves as a `black potential' for unidirectional as well 
as bidirectional incidence. We demonstrate again a non-Hermitian asymmetric Hulthen potential 
in Fig. 5(a) to show the almost vanishing scattering coefficients (Fig. 5(b, c, d)) for bidirectional incidence.
\\

\includegraphics[scale=1.2]{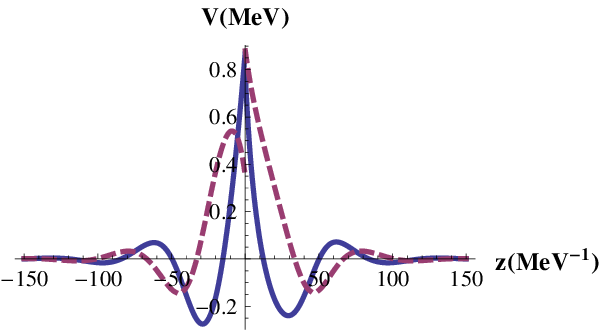} \ a \ \ \includegraphics[scale=1.3]{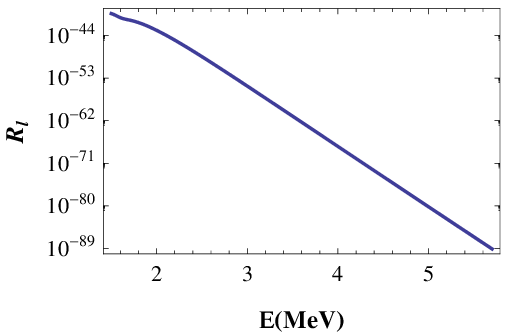} \ b \ \\
\includegraphics[scale=1.4]{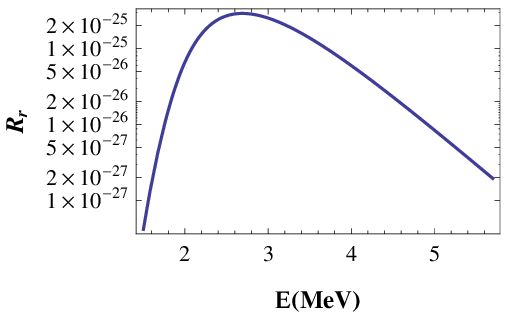} \ c \ \ \includegraphics[scale=1.4]{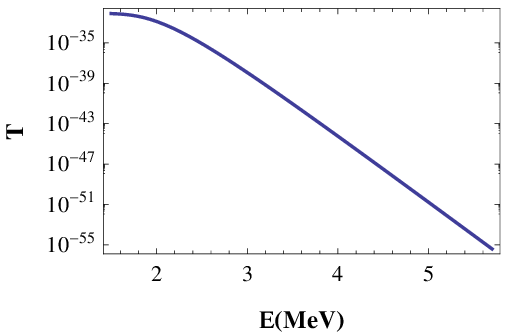} \ d \ \\

{\it Fig. 5: Asymmetric Hulthen potential is shown in (a) ($q_1 =0.2, q_2 = 0.4, v_1= v_2 = 1 MeV, a_1 = a_2 = 0.1 MeV, \phi_{1} = 1.8, \phi_{2} = 1.0, \gamma_{1} = \gamma_{2} = 2.0, \beta_{1} =\beta_{2} = 0.2,$) which leads to CPA, the critical values
of $\gamma_{1,2}$ are $\gamma_{1}^c =2.3$ and $ \gamma_{2}^c = 2.04$. The 
behaviors of scattering coefficients $R_l,R_r$ and $T (=T_l=T_r)$
are shown in Figs. (b,c,d) which satisfy bidirectional null scattering.}   \\

\hspace{-0.3in} This potential absorbs the energies of relativistic particles even in the 
ultrahigh range. Similarly all the different shapes of the potential in Fig. 3
also lie into the parametric regime of broadband CPA.  
The absorption rate increases with incidence energy which is an astonishing result for
this potential.

\hspace{0.3in} To explain this result we wish to put light on three dimensional scattering by an object 
where the total scattering cross section $\sigma_{tot}$ is 
the summation of total elastic scattering cross section $\sigma_{el}$ and the total 
cross section $\sigma_{abs}$ due to absorption plus inelastic scattering \cite{sch}. A 
perfect absorber is an object for which, 
\beq
\sigma_{el}\approx \sigma_{abs}\approx A_{tot} \ , \ \ \mbox{so that } \ \ 
\sigma_{tot}\approx 2 A_{tot}
\eeq
where  $A_{tot}$ is the cross-sectional area of the object. In Ref. \cite{sch} the perfect
absorption is discussed for non-relativistic case by Eikonal approximation ($V<<E)$.  
Following the same discussion, the scattering cross-sections for KG equation (in natural unit
with $m=1$) for a complex potential $V(r)=V_R(r)\pm i V_I(r)$ (where $r=r(x,y,z)$) is written as,  
\bea
\sigma_{el}&\approx & \int_{-\infty}^{+\infty}\int_{-\infty}^{+\infty}\left\{1+ e^{2\theta} -2 \cos\left[\frac{1}{2k}\int_{-\infty}^{+\infty}({V_R}^2-{V_I}^2)dz-(1+
\frac{1}{2 k^2})
\int_{-\infty}^{+\infty}V_R dz\right] e^{\theta}\right\} dx dy ; \nonumber \\
&&
\label{el}
\eea
and 
\beq
\sigma_{abs}\approx  \int_{-\infty}^{+\infty}\int_{-\infty}^{+\infty}\left(1- e^{2\theta}\right) dx dy
\label{abs}
\eeq
where
\beq
\theta =\mp 
\frac{2}{k}\int_{-\infty}^{+\infty}V_RV_I dz \pm2(1+\frac{1}{2 k^2})\int_{-\infty}^{+\infty}
V_Idz \ \mbox{with} \ k=\sqrt{E^2-1}
\label{th}
\eeq
For a step like potential as in our case we define quantities $\theta_I, \theta_R$ as,
\bea
\theta_I\equiv \int_{-\infty}^{+\infty} V_I(x,y,z) dz &=&\int_{-\infty}^{0} V_I^-(x,y,z) dz+
\int_{0}^{+\infty} V_I^+(x,y,z) dz \ ; \nonumber \\
\theta_R\equiv \int_{-\infty}^{+\infty} V_R(x,y,z) dz &=&\int_{-\infty}^{0} V_R^-(x,y,z) dz+
\int_{0}^{+\infty} V_R^+(x,y,z) dz \ .
\eea
We apply this approach to the Hulthen potential in Eq. \ref{hul1}. The imaginary parts 
of this complexified potential for $z<0$ and $z>0$ are written in terms of the parameters
in Eq. \ref{para} as,
\bea
V_I^-(r)&=&\frac{e^{a_1 r \cos \gamma_1}{v_1} \Big\{e^{a_1 r \cos \gamma_1} q_1 \sin\phi_1+\sin (a_1 r \sin \gamma_1)\Big\}}
{1+q_1^2 e^{2 a_1 r \cos \gamma_1} -2 q_1 e^{a_1 r \cos \gamma_1} \cos (\phi_1+a_1 r \sin \gamma_1)} \nonumber \\
V_I^+(r)&=&\frac{e^{-a_2 r \cos \gamma_2}{v_2} \Big\{e^{-a_2 r \cos \gamma_2} q_2 \sin\phi_2-\sin (a_2 r \sin \gamma_2)\Big\}}
{1+q_2^2 e^{-2 a_2 r \cos \gamma_2} -2 q_2 e^{-a_2 r \cos \gamma_2} \cos (\phi_2-a_2 r \sin \gamma_2)}
\label{vmvp}
\eea
Now there are many possibilities for 
$\sigma_{el}\approx \sigma_{abs}$. However we are only interested in the case of higher energy ranges ($k>>1$) for which the integration in the expression $\theta$ in Eq. \ref{th} is reduced to ,
\beq
\theta_0 \equiv \lim_{k\rightarrow \infty}\theta= \pm 2\int_{-\infty}^{+\infty}
V_Idz \equiv \pm 2\theta_I
\eeq
Therefore at very high energy range the elastic and inelastic scattering cross-sections are 
independent of incidence energy and are written as,
\bea
\sigma_{el}&\approx & \int_{-\infty}^{+\infty}\int_{-\infty}^{+\infty}\left(1+ e^{2\theta_0} -2
\cos (\theta_R) e^{\theta_0}\right) dx dy ;
\label{el2}
\eea
\beq
\sigma_{abs}\approx  \int_{-\infty}^{+\infty}\int_{-\infty}^{+\infty}\left(1- e^{2\theta_0}
\right) dx dy
\label{abs2}
\eeq
The expressions in Eqs. \ref{el2} and \ref{abs2} show that at very higher energy regions 
the scattering cross-sections are getting saturated (i.e. depend on the potential parameters
only).
For $\theta_0\rightarrow 0, \ \sigma_{abs}\rightarrow 0$ explains absorption less scattering
due to a real potential. But $\theta_0\rightarrow -\infty$ is also a possible case even though 
$V_I<<E$ and the exponentials become negligibly small compared to unity for all trajectories 
x, y that pass through the object. In this case  
$\sigma_{el}\approx \sigma_{abs}\approx A_{tot}$ which defines a total 
absorptivity for the higher energy range. This justifies our result, saturation of 
absorptivity towards unity for higher energy range.

\section {Waveguide analog of non-Hermitian Hulthen potential}
Waveguide analogy \cite{wg} of a quantum mechanical potential is very useful to study the 
problem in  classical regime. In this section we prescribe the
cross-sectional view of the waveguide which is analogous to this relativistic non-Hermitian 
Hulthen potential. We start from the KG equation for a free particle,
\beq
\frac{d^2\phi(z)}{dz^2}+\Big\{\frac{E^2}{c^2\hbar^2}-\frac{m^2c^2}{\hbar^2}\Big\}
\phi(z)\equiv\frac{d^2\phi(z)}{dz^2}+k_g^2
\phi(z)=0
\label{kg2}
\eeq
with the quantum mechanical propagation constant $k_g$ which is written in terms of 
de-Broglie wavelength as,
\beq
k_g=\frac{2 \pi}{\lambda_E} \ , \ \mbox{where} \ \lambda_E=\frac{2 \pi \hbar c}{\sqrt{E^2-m^2c^4}}
\eeq 
The propagation constant of a waveguide is a function of the geometrical structure of the
waveguide, its analogy with the quantum mechanical propagation constant offers a physical insight of the mathematical description of quantum scattering
through a potential. To understand further in a clear way we draw the geometrical 
view of a waveguide which is analogous to a general step potential $U(z)$ [Fig. 6(a)], \\

\begin{center}
\ \ \includegraphics[scale=1.0]{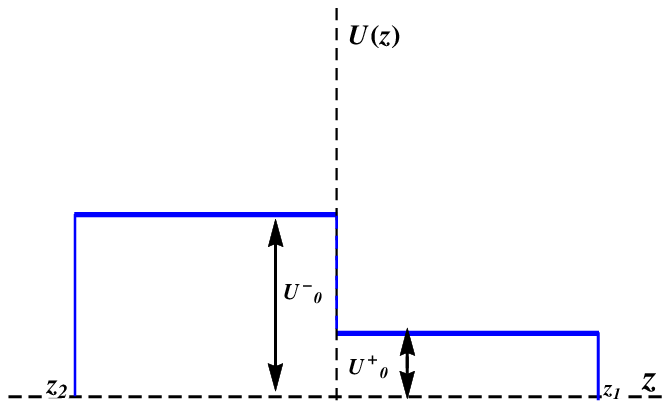} (a) 
\includegraphics[scale=1.0]{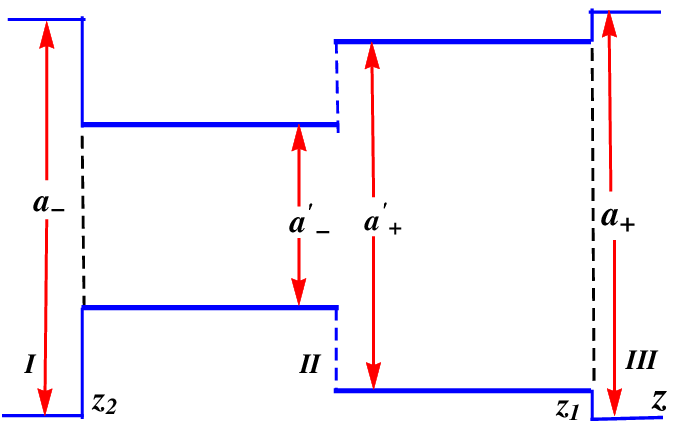} (b) \\
\end{center}
{\it Fig. 6: The figure in right (Fig. 6(b)) demonstrates the geometrical view of a
waveguide which is analogous to the general step potential shown in the left (Fig. 6(a)). Cross-sectional length of the waveguide is 
plotted vertically whereas $a_\pm(z)$, the cross-sectional distribution is plotted horizontally.  } \\

Eq. \ref{kg2} is now compared to an electromagnetic wave equation \footnote{For 
mathematical simplicity we select
the $TE_{10}$ mode of the plane electromagnetic wave \cite{wg} whose electric and magnetic field components are oriented in a such a way that $E_x=H_y=E_z=0$. The time-independent electromagnetic wave equation for this $TE_{10}$ wave propagation through
a waveguide is written in Eq. \ref{wg} for the region I and III of Fig. 6(b).},
\beq
\frac{d^2 H_z}{dz^2}+k_{wg}^2H_z=0, \ \ \mbox {with} \ \ k_{wg}=\frac{2\pi}{\lambda}\left[1-(\frac{\lambda}{2 a_\pm})^2\right]^{1/2}; 
\label{wg}
\eeq
for the waveguide in the region I and III of Fig. 6(b).
$a_\mp$ are the cross-sectional lengths of the waveguide in the region 
I and III as depicted in Fig. 6(b). The waveguide has a cutoff wavelength $\lambda_c=2a_\mp$ for  
the incoming waves from region I and region III respectively.
By the analogy between $k_g$ and the electromagnetic propagation constant $k_{wg}$ we write,
\beq
\frac{2 \pi}{\lambda_E}\propto \frac{2\pi}{\lambda}\left[1-(\frac{\lambda}{2 a_\pm})^2\right]^{1/2}=\frac{2\pi s}{\lambda}\left[1-(\frac{\lambda}{2 a_\pm})^2\right]^{1/2}, \ ;
\label{v0}
\eeq
where $s$ is the constant of proportionality between the two propagation constants. Now for
$U(z)\neq0$ the quantum mechanical propagation constant ($k_{gs}$) for scattering  of spin less particle due to the general step potential is calculated using the KG-equation in Eq. \ref{kg} as,
\bea
k_{gs}&=&\left[\frac{\left\{E-U(z)\right\}^2}{c^2\hbar^2}-\frac{m^2c^2}{\hbar^2}\right]^{1/2}
= 2\pi\left[\frac{E^2-m^2c^4}{4 \pi^2\hbar^2 c^2}\left(1-\frac{2 E U(z)-U^2(z)}{E^2-m^2c^4}\right)\right]^{1/2} \nonumber \\
&=&\frac{2\pi}{\lambda_E}\left[1-\frac{2 E U(z)-U^2(z)}{E^2-m^2c^4}\right]^{1/2}
\eea
So the analogy between KG-equation for scattering due to the general step potential and the electromagnetic-wave equation 
for $\lambda < {2a'}_\pm$ (where ${a'}_\pm$ are the cross-sectional lengths of the waveguide corresponds to the peaks of the potential $U(z)$ (i.e. $U^\pm _0$)) in the region II of the Fig. 6 is written as, 
\beq
\frac{2 \pi}{\lambda_E}\left[1-\frac{U(z)(2 E-U(z))}{E^2-m^2c^4}\right ]^{1/2}=\frac{2\pi s}{\lambda}\left[1-(\frac{\lambda}{2 a_\pm(z)})^2\right ]^{1/2}
\label{scatt}
\eeq
Here $a_\pm(z)$ is the cross sectional distribution of the waveguide along $z$-direction 
for $z>0$ and $z<0$ respectively in region II (depicted in Fig. 6(b)). Note $a_\pm(z<z_2)=a_+$
and $a_\pm(z>z_1)=a_-$. We can see from Fig. 6(b) (as $U_0^->U_0^+$) that the relation between ${a'}_+$ and ${a'}_-$ can be written as,
\beq
a'_{+}=a'_{-}+2\left (U_0^--U_0^+ \right ) \ \ \mbox{with} \ \ a'_{-}>0.
\label{rel1}
\eeq
Now by dividing Eq. (\ref{scatt}) by Eq. (\ref{v0}) 
and squaring we get,
\beq
\frac{E^2-m^2c^4}{U(z)(2 E-U(z))}=\frac{(\frac{2a_\pm}{\lambda })^2-1}{(\frac{a_\pm}{a_\pm(z)})^2-1}
\label{az}
\eeq  
The step potential in Fig. 6(a) can be written as $U^\pm (z)=U_0^\pm g(z)$, where $g(z)=0$ for $z> z_1$
and $z< z_2$ and $g(z)=(\theta(-z)+\theta(z))$ for $z_2\leq z\leq z_1$ (where $\theta(\pm z)$ is defined in Eq. \ref{thet}). Now to calculate waveguide 
cross-section corresponding to this potential we consider the limit, incident particle energy $E\rightarrow 
U_0^\pm$ which is equivalent to  $\lambda \rightarrow 2 a'_{\pm}$ for case of waveguide. Putting
this limiting condition in Eq. (\ref{az}) we get the waveguide cross-sections as a function of $z$ as,
\beq
a_{\pm}(z)=a_{\pm}\left [1+G_{\pm}(z)\left \{\left (\frac{a_{\pm}}{a'_{\pm}}\right )^2-1\right \}\right ]^{-1/2}, \ \mbox{where} \ \ G_{\pm}(z)=\frac{(2U_0^\pm-U_0^\pm g (z))U_0^\pm g (z)}{(U_0^\pm )^2-m^2c^4}
\label{cr1}
\eeq
This expression of waveguide cross-section (for the potential in Fig. 6(a)) implies that at $z< z_2$
and $z> z_1 \ $  $G_{\pm}(z)=0$ and hence the cross-sections are 
$a_{\pm}(z)=a_{\pm}$ as depicted in the Fig. 6(b). At region II (i.e. $z_2<z<z_1$) 
$g(z)=1$ and hence $G_{\pm}(z)=\frac{1}{1-\left (mc^2/U_0^\pm\right )^2}$ leads to
the minimum cross-sectional lengths inside the waveguide,
\beq
a^{min}_{\pm}=a_{\pm}\left [1+\left (\frac{1}{1-\left (mc^2/U_0^\pm\right )^2}\right )\left \{\left (\frac{a_{\pm}}{a'_{\pm}}\right )^2-1\right \}\right ]^{-1/2},
\eeq
Therefore this waveguide 
corresponds to the scattering (due to this general step potential $U(z)$) of a spin less particle of incident wavelength $\lambda < 2 \times $ (least between $ a^{min}_{\pm}$).

Now in our case of Hulthen potential $V(z)$ (Eq. \ref{hul1}) the cross-sectional distribution 
of the analogous waveguide simply can be understood from Eq. \ref{cr1}. The function
$G(z)$ for the general step potential is now replaced by $H(z)$ such that $V^\pm (z)=V_0^\pm 
H^\pm(z)$, where $V_0^\pm$ are the peaks of the potential $V (z)$ for $z\rightarrow 0^\pm$ 
respectively. 
For the Hulthen potential $V^\pm (z)$ are written as,
\bea
 V^+(z)&=&\frac{V_2}{e^{\alpha _2  z}-Q_2}=V_0^+\left(\frac{1-Q_2}{e^{\alpha _2  z}-Q_2}\right)\equiv V_0^+h^+(z), \ \ \mbox{where} \ \ V_0^+=\frac{V_2}{1-Q_2} \ ; \nonumber \\
 V^-(z)&=&\frac{V_1}{e^{-\alpha _1  z}-Q_1}=V_0^-\left(\frac{1-Q_1}{e^{-\alpha _1  z}-Q_1}\right)\equiv V_0^-h^-(z), \ \ \mbox{where} \ \ V_0^-=\frac{V_1}{1-Q_1} \ .
 \label{vv12}
\eea 
For the complex Hulthen potential $V(z)$ with the
parameters given in Eq. \ref{para} we consider a system of two single-mode
waveguides coupled \cite{dr, ga} to each other corresponding to the real and imaginary part of $V(z)$. We denote the two waveguides as $WG^R$ and $WG^I$ which are corresponding to the
real and the imaginary parts of $V(z)$ respectively. For a non-zero coupling between $WG^R$ and $WG^I$
the system mimics the quantum mechanical scattering through a non-Hermitian Hulthen 
potential. The cross-sectional distributions (analogous to Eq. \ref{cr1}) for the waveguides $WG^R$ and $WG^I$ are given as,
\bea
A^{R}_{\pm}(z)&=&A^{R}_{\pm}\left [1+Re[H_{\pm}(z)]\left \{\left (\frac{A^R_{\pm}}{A'^R_{\pm}}\right )^2-1\right \}\right ]^{-1/2}, \nonumber \\
A^{I}_{\pm}(z)&=&A^{I}_{\pm}\left [1+Im[H_{\pm}(z)]\left \{\left (\frac{A^I_{\pm}}{A'^I_{\pm}}\right )^2-1\right \}\right ]^{-1/2} 
\label{c1}
\eea
where
\beq
H^\pm(z)=\frac{(2V_0^\pm-V_0^\pm h^\pm (z))V_0^\pm h^\pm (z)}{(V_0^\pm )^2-m^2c^4}
\label{c2}
\eeq
$A^{R}_{\pm}$ and $A^{I}_{\pm}$ are the cross-sectional lengths of the $WG^R$ and $WG^I$ 
respectively when $Re[H_{\pm}(z)]=0$ and $Im[H_{\pm}(z)]=0$, i.e. at the region where $ 
V^\pm(z)=0$. $A'^R_{\pm}$ and  $A'^I_{\pm}$ are the cross-sectional lengths of the waveguides 
$WG^R$ and $WG^I$ respectively at $z\rightarrow 0^\pm$ which actually correspond to the peaks 
(i.e. $V^\pm _0$) of the potential $V(z)$. Relation between $A'^R_{\pm}$, $A'^R_{\pm}$ and 
$A'^I_{\pm}$, $A'^I_{\pm}$ (assuming $Re[V_0^-]>Re[V_0^+]$ and $Im[V_0^-]>Im[V_0^+]$) are written as in Eq. \ref{rel1}, 
\bea
A'^R_{+}&=&A'^R_{-}+2\left (Re[V_0^-]-Re[V_0^+] \right ) \ \mbox{with} \ A'^R_{-}>0 ,\ \nonumber \\
A'^I_{+}&=&A'^I_{-}+2\left (Im[V_0^-]-Im[V_0^+] \right ) \ \mbox{with} \ A'^I_{-}>0.
\label{rel2}
\eea
From  Eq. \ref{vv12} we see that $h^\pm (z)=1$ for $z\rightarrow 0^\pm$, 
thus the minimum cross-sectional lengths of $WG^R$ and $WG^I$
are written from Eqs. \ref{c1} and \ref{c2} as,
\bea
{A^R}^{min}_{\pm}&=&A^{R}_{\pm}\left [1+\left (\frac{1}{1-\left (mc^2/Re[V_0^\pm]\right )^2}\right )\left \{\left (\frac{A^R_{\pm}}{A'^R_{\pm}}\right )^2-1\right \}\right ]^{-1/2}, \nonumber \\
{A^I}^{min}_{\pm}&=&A^{I}{\pm}\left [1+\left (\frac{1}{1-\left (mc^2/Im[V_0^\pm]\right )^2}\right )\left \{\left (\frac{A^I_{\pm}}{A'^I_{\pm}}\right )^2-1\right \}\right ]^{-1/2}
\label{mn}
\eea
The electromagnetic wave propagating through the system of coupled waveguide experience both 
elastic scattering (due to $WG^R$) and inelastic scattering (due to $WG^I$). This coupled
waveguide describes the scattering of a spin less particle due to non-Hermitian Hulthen 
potential if the wavelength of the incoming wave is smaller than the two times of the least 
cross-sectional lengths among the values of ${A^R}^{min}_{\pm}$ and ${A^I}^{min}_{\pm}$.
By taking the cross-sectional lengths of the two ends of waveguides $WG^R$ and $WG^I$ as  
$A^R_{\pm}=A'^R_{-}+2 Re[V_0^-]$ and $A^I_{\pm}=A'^I_{-}+2 Im[V_0^-]$  respectively we present
the geometrical views of $WG^R$ and $WG^I$ in Fig. 7 (b, d) corresponding to the parameters of the real and imaginary parts of the potential in 
Fig. 7(a, c) which leads to broadband CPA.
The four cross-sectional lengths inside the waveguides in Fig. 7 (b, d) are calculated from Eq. \ref{mn} as 
${A^R}^{min}_{+}=197.026 \ nm$ and ${A^R}^{min}_{-}=197 \ nm={A^I}^{min}_{-}$
and ${A^I}^{min}_{+}=197.01 \ nm$. Therefore
for these two coupled waveguides in Fig 7(b, d) the operating wavelength of the incoming wave
must be smaller than two times of ${A^R}^{min}_{-}$ i.e. $\lambda_{max} \approx  394 \ nm $.
Similar waveguides with different physical dimension can also be constructed for different parametric regions of the 
non-Hermitian Hulthen potential which leads to broadband CPA.

\begin{center}
\includegraphics[scale=1.2]{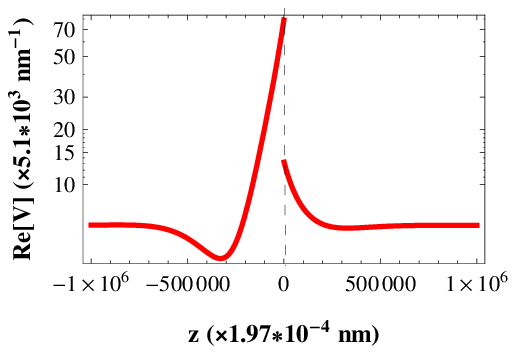} a  \includegraphics[scale=1.22]{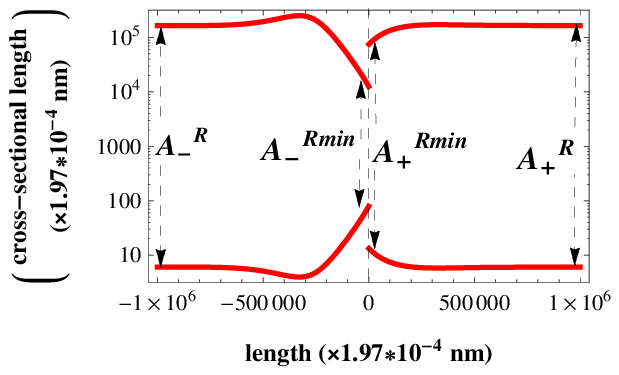} \ b \\
\includegraphics[scale=1.2]{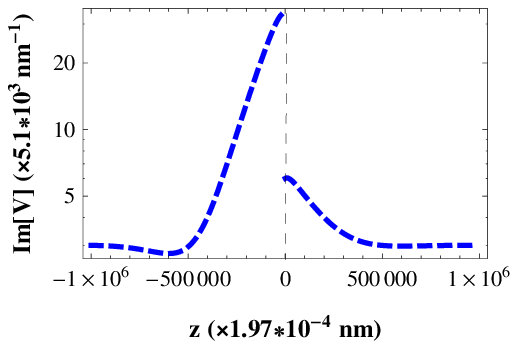} c  \includegraphics[scale=1.23]{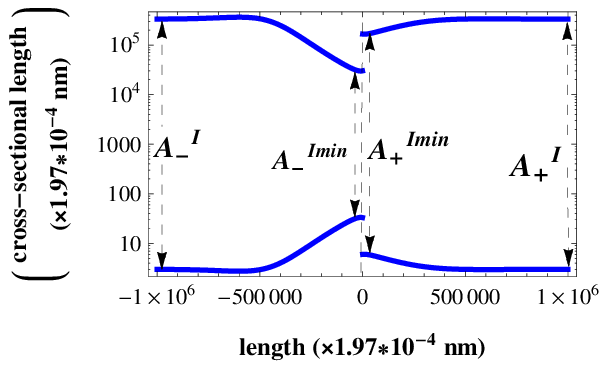} \ d \\
\end{center}
{\it Fig. 7: Real and imaginary parts of a non-Hermitian Hulthen potential is plotted in 7(a) 
and 7(c) respectively with the parameters $q_1 = q_2 = 0.4, v_1= 2.53\times 10^{5} \ nm^{-1}, v_2 = 2.53\times 10^{4} \ nm^{-1}, a_1 =a_2 = .051 \ nm^{-1}, \phi_{1} = \phi_{2} = 0.3, \gamma_{1} = \gamma_{2} = 2.5, 
\beta_{1} =\beta_{2} = 0.2,$}. The cross-sectional lengths and dimension of the corresponding
waveguides are shown in 7(b) and 7(d).   \\

\section{Conclusions and Discussions}
Null scattering (CC, CPA) and super-scattering (lasing), two extreme scattering situations are important tools in characterizing scattering due to complex potentials. In particular the occurrence of total
absorption through a complex potential becomes very exciting due to the recent discovery of anti-laser 
which has number of technological applications. The occurrence of CPA and CC has so far been shown only
for certain discrete values of the incidence energies of the particle and that
too only for certain specific potentials. In a recent work we were able to show CPA for certain small ranges of incidence energies for well known nuclear potential \cite{hsn}. In this work we parametrize the
well known Hulthen potential in such a manner that it shows both unidirectional and bidirectional 
perfect absorption almost for the entire range of incidence energy. We have arrived to our conclusion analytically as well as graphically by considering both unidirectional and bidirectional incidence of the 
particles. The low energy scattering behavior of this potential is different from that of at high energies.
We have shown that further fine tuning is required within the ranges of the parameters to achieve CPA for
the low energy scattering. We would like to emphasize that over this entire range of incident energy the values of reflectance (R) and transmittance (T) are very very small ($\sim 10^{-24}$ or less). We call this 
potential as a `black potential' which leads to broadband CPA. The values of R and T become less and less as we go on increasing incident energy.
 To classically design and to test
our results one needs to know the waveguide analogy of this particular potential. We have 
provided the cross-sectional views of the waveguides analogous to our complex potential.  
Physical dimensions like lengths and cross-sectional lengths of the
waveguides corresponding to real and imaginary parts of the non-Hermitian Hulthen potential 
are demonstrated graphically and the operating wavelength ($\lambda_{max} \approx 394 \ nm $) of the coupled waveguides is analytically calculated. \\

{\it Acknowledgment:} BPM acknowledges the financial support from the Department 
of Science \& Technology (DST), Govt. of India, under SERC project sanction grant No. 
SR/S2/HEP-0009/2012 and MH is thankful to Dr. S. K. Shivakumar, Director, ISAC for his 
support to carry out this research work. AG acknowledges the Council of Scientific \& Indus-
trial Research (CSIR), India for Senior Research Fellowship.

\newpage
\appendix
\section{Appendix: Relativistic Scattering coefficients of Hulthen potential} 
\label{AppA}
The general solutions of KG equation (by adopting natural units with $m=\hbar=c=1$ ) for the Hulthen potential at $z<0$ (i.e.
$V^-(z)=\frac{V_1}{e^{-\alpha_1 z}-Q_1}$)
are in terms of hyper-geometric function as follows,
\bea
\phi_L(z)&=&(1-Q_1e^{\alpha_1  z})^{\lambda_1} \Big\{A (Q_1e^{\alpha_1  z})^{\mu_1}  F(\mu_1-\nu_1+\lambda_1,\mu_1-\nu_1+\lambda_1, 1+2\mu_1;Q_1e^{\alpha_1  z}) \nonumber \\
&+& B (Q_1e^{\alpha_1  z})^{-\mu_1} F(-\mu_1-\nu_1+\lambda_1,-\mu_1+\nu_1+\lambda_1, 1-2\mu_1;Q_1e^{\alpha_1  z})\Big\}
\eea

Similarly the solutions for $V^+(z)=\frac{V_2}{e^{\alpha_2 z}-Q_2}$ (at $z>0$) are,
\bea
\phi_R(z)&=&(1-Q_2e^{-\alpha_2  z})^{\lambda_2}\Big\{C (Q_2e^{-\alpha_2 z})^{\mu_2} F(\mu_2-\nu_2+\lambda_2,\mu_2-\nu_2+\lambda_2, 1+2\mu_2;Q_2e^{-\alpha_2  z}) \nonumber \\
&+& H (Qe^{-\alpha  z})^{-\mu_2} F(-\mu_2-\nu_2+\lambda_2,-\mu_2+\nu_2+\lambda_2, 1-2\mu_2;Q_2e^{-\alpha_2  z})
\eea
where the following notations are used,
\beq
\mu_1 =ik/\alpha_1 ; \ \mu_2 =ik/\alpha_2; \ k=\sqrt{E^2-1} \ 
\label{mua}
\eeq
\beq
\lambda_1 =1/2+1/2\sqrt{1-\Big(\frac{2V_1}{\alpha_1 
Q_1}\Big)^2} \ ; \lambda_2 =1/2+1/2\sqrt{1-\Big(\frac{2V_2}{\alpha_2
Q_2}\Big)^2}
\label{l1l2a}
\eeq
\beq
\nu_1 =\sqrt{\mu_1 ^2+\lambda_1^2-\lambda_1 -\frac{2EV_1}{\alpha_1 ^2 Q_1}} ; \ \nu_2 
=\sqrt{\mu_2^2+\lambda_2^2-\lambda_2 -\frac{2EV_2}{\alpha_2^2 Q_2}}
\label{nu1nu2a}
\eeq
The asymptotic form of $\phi_L$ and $\phi_R$ are given as,
\bea
\phi_L (z\rightarrow -\infty ) & \rightarrow & A Q_1^{ik/\alpha _1}e^{ikz}+ B Q_1^{-ik/\alpha _1}e^{-ikz} \ ; \nonumber \\
\phi_R(z\rightarrow +\infty ) & \rightarrow & C Q_2^{ik/\alpha _2}e^{-ikz}+ H Q_2^{-ik/\alpha _2}e^{ikz}
\label{assymp}
\eea
Using Eq. \ref{assymp} the left and right handed scattering amplitudes are calculated as,
\beq
r_{l}=B/A=\frac{Q_1^{2\mu_1}\Big[-F_1F_2(N_1+N_2)+c_1F_1^{'}F_2+c_3F_2^{'}F_1\Big]}
{D}\equiv Q_1^{2\mu_1} \tilde{F_{rl}}
\label{rla}
\eeq
\beq
t_{l}=H/A=\frac{Q_1^{2\mu_1}\Big[2MF_1F_4+c_1F_1^{'}F_4-c_2F_4^{'}F_1\Big]}
{D}\left(\frac{1}{u}\right)\equiv  \frac{Q_1^{2\mu_1}}{u}\tilde{F_{tl}}
\label{tla}
\eeq

\beq
r_{r}=H/C=\frac{Q_2^{2\mu_2 
}\Big[-F_3F_4(N_1+N_2)+c_4F_3^{'}F_4+c_2F_4^{'}F_3\Big]}
{D}\equiv Q_2^{2\mu_2} \tilde{F_{rr}}
\label{rta}
\eeq

\beq
t_{r}=B/C=\frac{Q_2^{2\mu_2 }\Big[2MF_3F_2+c_4F_3^{'}F_2-c_3F_2^{'}F_3\Big] u}
{D}\equiv {Q_2^{2\mu_2}}{u}\tilde{F_{tr}}
\label{tta}
\eeq
where 
\beq
D=2MF_4F_2+F_4F_2(N_1+N_2)-c_2F_4^{'}F_2-c_3F_2^{'}F_4 \ \ \mbox{with} M=ik \ ;
\label{d}
\eeq
with 
\beq
N_1=\frac{Q_1\alpha_1 \lambda_1 }{1-Q_1} \ ; \ 
N_2=\frac{Q_2\alpha_2 \lambda_2 }{1-Q_2} \ ; 
\label{nna}
\eeq
\beq
 u=\frac{Q_2^{-\mu_2}(1-Q_2)^{\lambda_2}}{Q_1^{-\mu_1}(1-Q_1)^{\lambda_1}}
 \eeq
 \bea
c_1&=&\frac{(\mu_1 +\lambda_1 )^2-\nu_1 ^2}{1+2\mu_1 }(Q_1\alpha_1 ) \ ; \ c_2=\frac{(-\mu_1 +\lambda_1 )^2-\nu_1 ^2}{1-2\mu_1 
}(Q_1\alpha_1 ) \ ; \nonumber \\ 
c_3&=&\frac{(-\mu_2 +\lambda_2 )^2-{\nu_2} ^2}{1-2\mu_2 }(Q_2\alpha_2 ) \ ; \ c_4=\frac{(\mu_2 
+\lambda_2 )^2-{\nu_2} ^2}{1+2\mu_2 }(Q_2\alpha_2 ) \ .
\label{cssa}
\eea
The hyper-geometric functions are written as follows,
{\small\bea
\label{hyp1}
F_1&=&F(\mu_1 -\nu_1 +\lambda_1 , \mu_1 +\nu_1 +\lambda_1 , 1+2\mu_1 ; Q_1)\ ; \nonumber \\ F_1^{'}&=&F(\mu_1 -\nu_1 +\lambda_1 +1 , \mu_1 +\nu_1 +\lambda_1 +1 , 2+2\mu_1 ; Q_1) \ ; \\ 
\label{hyp2} \nonumber \\
F_2&=&F(-\mu_2 -\nu_2 +\lambda_2 , -\mu_2 +\nu_2 +\lambda_2 , 1-2\mu_2 ; Q_2) \ ; \ \nonumber \\ 
F_2^{'}&=&F(-\mu_2 -\nu_2 +\lambda_2 +1, -\mu_2 +\nu_2 +\lambda_2 +1,2-2\mu_2 ; Q_2) \\  \nonumber \\
F_3&=&F(\mu_2 -\nu_2 +\lambda_2 , \mu_2+\nu_2 +\lambda_2 , 1+2\mu_2 ; Q_2) \ ; \nonumber \\
\ F_3^{'}&=&F(\mu_2 -\nu_2 +\lambda_2 +1 , \mu_2 +\nu_2 +\lambda_2 +1 , 2+2\mu_2 ; Q_2) \ ; \\ \nonumber \\
F_4&=&F(-\mu_1 -\nu_1 +\lambda_1 , -\mu_1 +\nu_1 +\lambda_1 , 1-2\mu_1 ; Q_1) \ ; \nonumber \\
F_4^{'}&=&F(-\mu_1 -\nu_1 +\lambda_1 +1, -\mu_1 +\nu_1 +\lambda_1 +1, 2-2\mu_1 ; Q_1) ; 
\label{hyp4}
\eea}


\begin{thebibliography}{99}

\bibitem{ben4} C. M. Bender and S. Boettcher, {\em Phys. Rev. Lett.} {\bf 80}, 5243 (1998).
\bibitem{mos} A. Mostafazadeh, {\em Int. J. Geom. Meth. Mod. Phys.} {\bf 7}, 1191(2010) and references therein.
\bibitem{benr} C.M. Bender, {\em Rep. Progr. Phys.} {\bf 70} (2007) 947 and references therein.


\bibitem{opt1} Z. H. Musslimani, K. G. Makris, R. El-Ganainy, and D. N. Christodoulides, {\em Phys. Rev. Lett.} {\bf 100}, 030402 (2008).

\bibitem{opt2} C. E. Ruter, K. G. Makris, R. El-Ganainy, D. N. Christodoulides, M. Segev, D. Kip, {\em Nature Phys.} {\bf 6} 192, (2010); 

\bibitem{opt3}  R. El-Ganainy, K. G. Makris, D. N. Christodoulides and Z. H. Musslimani, {\em Opt. Lett.} {\bf 32}, 2632 (2007).

\bibitem{eqv1} A. Guo et al, {\em Phys. Rev. Lett.} {\bf 103}, 093902 (2009).

\bibitem{ent} A. Ghatak and B. P. Mandal, {\em J. Phys. A: Math. Theor.}
{\bf 45}, 355301 (2012).

\bibitem{bmsm} B. P. Mandal and S S. Mahajan  {\em arXiv:1312.0757}, (2013).

\bibitem{ph} B. P. Mandal, B. K. Mourya, and R. K. Yadav (BHU),
{\em Phys. Lett. A} {\bf 377}, 1043 (2013).

\bibitem{bpm} B. P. Mandal, {\em Mod. Phys. Lett. A} {\bf 20}, 655(2005).

\bibitem{cal} B. P. Mandal and A. Ghatak, {\em J. Phys. A: Math. Theor.}
 {\bf 45}, 444022 (2012) .


\bibitem{ep0} T. Kato, {\em Perturbation Theroy of Linear Operators}, {\bf Springer}, Berlin, (1966).
\bibitem{ep1} M. V. Berry, {\em Czech. J. Phys.} {\bf 54}, 1039 (2004).
\bibitem{ep2} W. D. Heiss, {\em Phys. Rep.} {\bf 242}, 443 (1994).


\bibitem{ss1} A. Mostafazadeh, {\em Phys. Rev. Lett.} {\bf 102}, 220402 (2009).

\bibitem{ss2} A. Mostafazadeh, M. Sarisaman, {\em Phys. Lett. A} {\bf 375}, 3387 (2011).

\bibitem{aop} A. Ghatak, R. D. Ray Mandal, B. P. Mandal, {\em Ann. of Phys.} {\bf 336}, 540 (2013).

\bibitem{ss3} A. Ghatak, J. A. Nathan, B. P. Mandal, and Z. Ahmed, {\em J. Phys. A: Math. 
Theor.} {\bf 45},  465305 (2012).

\bibitem{inv2}  S. Longhi, {\em J. Phys. A: Math. Theor.} {\bf 44}, 485302 (2011).

\bibitem{inv1} A. Mostafazadeh, {\em Phys. Rev. A} {\bf 87}, 012103 (2013).

\bibitem{resc} L. Deak, T. Fulop, {\em Ann. of Phys.} {\bf 327}, 1050 (2012).

\bibitem{cpa00} C. F. Gmachl, {\em Nature} {\bf 467}, 37 (2010).

\bibitem{cpa01} S. Longhi, {\em Physics} {\bf 3}, 61(2010).

\bibitem{cpa011} W. Wan, Y. Chong, L. Ge, H. Noh, A. D. Stone, H. Cao, {\em Science} {\bf 331}, 889 (2011).

\bibitem{cpa02} N. Liu, M. Mesch, T. Weiss, M. Hentschel, and H. Giessen, {\em Nano Lett.} {\bf 10}, 2342 (2010).

\bibitem{cpa1} H. Noh, Y. Chong, A. Douglas Stone, and Hui Cao, {\em Phys. Rev. Lett.} {\bf 108}, 6805 (2011).

\bibitem{cpa0} A. Mostafazadeh and M. Sarisaman, {\em Proc. R. Soc. A} {\bf 468}, 3224 (2012).

\bibitem{cpa2} S. Longhi, {\em Phys. Rev. A} {\bf 83}, 055804 (2011).

\bibitem{cpa3} S. Dutta-Gupta, R. Deshmukh, A. Venu Gopal, O. J. F. Martin, and S.
Dutta Gupta, {\em Opt. Lett.} {\bf 37}, 4452 (2012).

\bibitem{cpa4} N. Liu, M. Mesch, T. Weiss, M. Hentschel, and H. Giessen, {\em Nano Lett.} {\bf 10} 2342 (2010).


\bibitem{cc0} M. Cai, O. Painter, and K. J. Vahala, {\em Phys. Rev. Lett. } {\bf 85}, 74 (2000).

\bibitem{cc1} J. R. Tischler, M. S. Bradley, and V. Bulovic, {\em Opt. Lett.} {\bf 31},
2045 (2006)

\bibitem{cc2} S. Dutta Gupta, {\em Opt. Lett.} {\bf 32}, 1483 (2007).

\bibitem{cc3} S. Balci, C. Kocabas, and A. Aydinli, {\em Opt. Lett.} {\bf 36}, 2770 (2011).

\bibitem{cc4} S. Balci, Er. Karademir, C. Kocabas, and A. Aydinli, {\em Opt. Lett.} {\bf 36}, 3041 (2011).

\bibitem{dp} A. Mostafazadeh,  {\em J. Phys. A: Math. Theor.} {\bf 45}, 444024 (2012).

\bibitem{hsn} M. Hasan, A. Ghatak, B. P. Mandal, {\em Ann. of Phys. } {\bf 344 },  17 (2014).


\bibitem{hul1} J. Guo, X. Fang, {\em Can. J. Phys.}, {\bf  87} 1021 (2009).

\bibitem{ref1} Pham Loi Vu, {\em Acta App. Math.}, {\bf 49} 107 (1997).

\bibitem{ref2} M S Marinov and B. Segev, {\em J. Phys. A: Math. Gen.} {\bf 29} 2839 (1996).
 
\bibitem{hbook}  I. S. Gradshteyn, I. M. Ryzhik, A. Jaffrey {\em 'Table of Integrals, Series and Products'}, {\it Academic press} (1965).
 
\bibitem{sch}  L. I. Schiff, {\em 'Quantum Mechanics'}, {\it McGraw-Hill international edition} (1968).

\bibitem{wg} M. Campi, {\em Am. J. Phys.}, {\bf 35}, 133 (1967).

\bibitem{dr} A. Paul, A. Saha, S. Bandopadhyay, and B. Dutta-Roy, {\em Eur. Phys. J. D} {\bf 42}, 495 (2007).

\bibitem{ga} G. S. Agarwal and K. Qu, {\em Phys. Rev. A } {\bf 85}, 031802 (R) (2012).

\end{thebibliography}
\end{document}